\documentclass{emulateapj}

\slugcomment{to appear in Astrophysical Journal}

\shorttitle{Multiplicity of Nova Solutions}
\shortauthors{Kato \& Hachisu}

\begin{document}


\title{MULTIPLICITY OF NOVA ENVELOPE SOLUTIONS AND OCCURRENCE OF 
OPTICALLY THICK WINDS
}


\author{Mariko Kato}
\affil{Department of Astronomy, Keio University, 
Hiyoshi 4-1-1, Kouhoku-ku, Yokohama 223-8521, Japan:}
\email{mariko@educ.cc.keio.ac.jp}

\and

\author{Izumi Hachisu}
\affil{Department of Earth Science and Astronomy, 
College of Arts and Sciences,
University of Tokyo, Komaba 3-8-1, Meguro-ku, Tokyo 153-8902, Japan;}
\email{hachisu@ea.c.u-tokyo.ac.jp}




\begin{abstract}
We revisit the occurrence condition of optically thick winds reported 
by \citet{kat85} and \citet{kat89} 
who examined mathematically nova envelope solutions with an old opacity  
and found that optically thick winds are accelerated only in massive white 
dwarfs (WDs) of $\gtrsim 0.9~M_{\odot}$. With the OPAL opacity we find 
that the optically thick wind occurs 
for $\gtrsim 0.6~M_{\odot}$ WDs and that the occurrence of winds depends 
not only on the WD mass but also on the ignition mass. 
When the ignition mass is larger than a critical value, winds are suppressed 
by a density-inversion layer. 
Such a static solution can be realized in WDs of mass 
$\sim 0.6-0.7~M_{\odot}$. We propose that sequences consisting only  
of static solutions correspond to 
slow evolutions in symbiotic novae like PU Vul because PU Vul 
shows no indication of 
strong winds in a long-lasted flat peak followed by a very slow decline 
in its light curve.
\end{abstract}





\section{INTRODUCTION } \label{sec_introduction}

Nova is a thermonuclear runaway event on a white dwarf (WD).
During a nova outburst, the envelope on the WD expands to a giant size.
It reaches the maximum brightness and then becomes dark as 
the envelope loses its mass due to strong winds.  
Evolutions of such shell flashes have been theoretically followed 
by many authors with hydrodynamic/hydrostatic time-dependent calculations \citep
{pac78,spa78,nar80,ibe82,pri86,pri95,sio79}, by using a sequence of 
steady-state solutions \citep{kat83b,kat88,kat94h},
and by using a sequence of static solutions \citep{fuj82,mac83}.

The time scale of nova evolution depends strongly on the WD mass. 
Massive WDs correspond to fast novae while less massive WDs 
do to slow novae. When the optically thick winds occur 
a large part of the envelope is blown off by winds, and the nova duration 
is drastically shortened \citep{kat94h,pri95}. 
In less massive WDs winds are relatively weak, and in some cases, no optically 
thick winds occur. However, the occurrence condition of winds
has not been well studied yet.

\citet{kat85} mathematically examined the occurrence of optically thick winds
with the Kramers opacity  and found that
the optically thick wind occurs when the luminosity approaches the Eddington 
luminosity and the opacity increases outward. \citet{kat89} showed that 
the optically thick wind occurs only in massive WDs of $M_{\rm WD} \geq 
0.9~M_{\odot}$, where $M_{\rm WD}$ is the WD mass.   
Wind acceleration is due to opacity enhancements at hydrogen and helium 
ionization zones. They also showed that a nova evolution can be well 
represented by a sequence consisting only of static solutions for 
$M_{\rm WD} < 0.9 ~M_{\odot}$, and by a sequence of wind and static 
solutions for $M_{\rm WD} \geq 0.9~M_{\odot}$.

In the beginning of 1990's, opacity tables are revised 
\citep[OPAL opacity:][and references therein]{igl96}, which show 
a prominent peak at $\log T $(K)$ \sim 5.2$ strong enough to accelerate 
winds even in less massive WDs of $M_{\rm WD}\sim 0.5-0.6~M_{\odot}$ 
\citep{kat94h}. However, the occurrence condition of winds has not been 
clarified yet, because the OPAL opacity has an intense peak at much higher 
temperature region than the helium ionization zone.  
Therefore, we re-examine the occurrence of optically thick winds 
for the OPAL opacity. We focus on less massive WDs in view of the 
application to slow novae or symbiotic novae in which less massive WDs
certainly inhabit.

In \S \ref{sec_model} we introduce our simplified model for nova outbursts.
Then we  present evolution sequences 
for three different WD masses in \S \ref{sec_novaevolution}. 
Temporal changes of internal structure during nova outbursts are shown 
in \S \ref{sec_internal}.  The 
occurrence condition of optically thick winds on various WD masses  
is presented in \S \ref{sec_variousWDmass}. 
Discussion and summary follow in \S\S  \ref{sec_discussion} and \ref{sec_summary}.

\section{A SIMPLIFIED MODEL}   \label{sec_model}

\begin{figure}
\epsscale{1.15}
\plotone{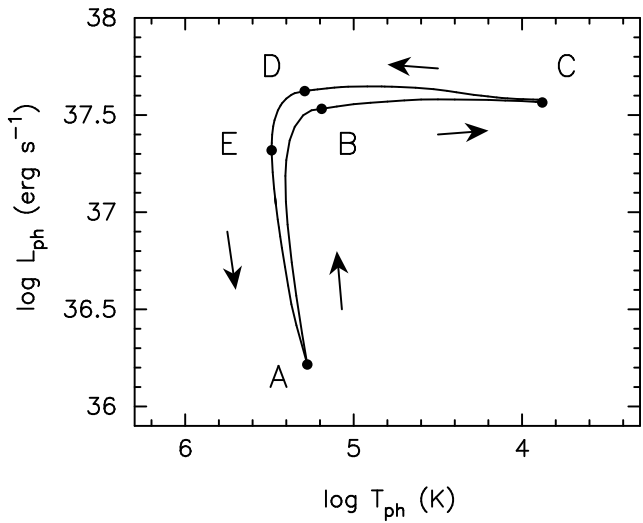}
\caption{
A schematic HR diagram of one cycle of shell flashes.
Thermonuclear runaway starts on an accreting WD at point A.
The star brightens up and its envelope begins to expand. 
After the star reaches the maximum expansion at point C the star moves
leftward along with the decreasing envelope mass sequence and 
the surface temperature increases with time. 
The nuclear burning extinguishes at point E and the star goes back to the 
original stage.
When the optically thick wind occurs, it begins at point B and ends 
at point D. 
\label{hr_diagram}
}
\end{figure}

Figure \ref{hr_diagram} presents a schematic HR diagram of one 
cycle of nova outbursts.
After thermonuclear runaway sets in on an accreting white dwarf (point A), 
the star brightens up and  its envelope expands to a giant size. 
Optically thick winds blow in massive WDs during the extended envelope stage
(from point B to point D through C).
The star reaches the maximum radius at point C, where the envelope 
settles down into a thermal equilibrium in which 
nuclear energy generation is balanced with radiative loss. Afterwards, the star
moves leftward keeping luminosity almost constant.
Hydrogen nuclear burning stops at point E, and the star cools down to point A.

When the wind occurs we solve the equations
of motion, mass continuity, radiative diffusion, and conservation of energy,
from the bottom of the hydrogen-rich envelope through the photosphere
assuming steady-state and spherical symmetry.
This steady-state is a good approximation in the decay phase or 
in weak shell flashes \citep{pri86,kat89sh}.
When the optically thick wind does not occur, we solve the 
equation of hydrostatic balance instead of the equation of motion. 
In the rising phase, we integrated energy conservation equation without 
energy generation term due to nuclear burning and later 
estimated energy generation 
using the temperature and density obtained  \citep{fuj82}.
In the decay phase we set the condition that the energy generation is 
balanced with radiative energy loss. These equations and method of 
calculations are already published in \citet{kat94h}.

In the rising phase, i.e., from point A to C in 
Figure \ref{hr_diagram}, we approximate the nova evolution by 
a sequence of envelope solutions as follows. 
When the optically thick wind occurs we use static solutions from point A to B 
and  steady-state wind solutions from point B to C in Figure \ref{hr_diagram}. 
When the  wind does not occur the entire sequence consists only of 
static solutions. 
The occurrence of optically thick winds is detected by the condition described 
in \citet{kat85} :(1) the photospheric luminosity approaches the Eddington limit 
and (2) at the same time the thermal energy at the photosphere is comparable
to the gravitational energy. 
We further assume that the envelope mass does not decrease much 
in the rising phase, i.e., we approximate the rising-phase 
by a sequence of envelope solutions with a constant mass which
we define as the ``ignition mass'', $\Delta M_{\rm ig}$.
This approximation may be too simple but we are interested in 
qualitative properties of the envelope evolution, which are 
hardly changed even if we assume decreasing envelope mass.

In the decay phase the envelope settles down into a thermal equilibrium, 
that is, the energy release due to nuclear burning is balanced to the outgoing 
energy flux.
We approximate this stage by a sequence of optically thick 
wind solutions (from point C to point D) and static solutions 
(from point D to point E).  When the wind does not occur,
the entire decay phase are approximated by a sequence only of static solutions. 
In this sequence (from point C to E) the envelope mass is decreasing 
from its initial value of $\Delta M_{\rm ig}$ at point C.
Time-evolution is calculated from the mass decreasing rates by nuclear burning and 
wind mass-loss if it occurs \citep[see][for more details]{kat94h}.  
Hydrogen burning stops at point E and 
after that, from point E to A, the evolution is followed by a static solution 
with a constant envelope mass.

We have neglected effects of convection in wind solutions, because convection 
is ineffective in a rapidly expanding envelope \citep{kat94h}.  
Convective energy transport is calculated in static solutions using 
the mixing length theory with $\alpha=1.5$ otherwise specified. Effects of 
the $\alpha$ parameter on our results are discussed in \S \ref{sec_mixinglength}. 
We use the OPAL opacity \citep{igl96}.

We further assume that the chemical composition of the envelope is uniform 
throughout the envelope and is unchanged with time, because the 
convection widely develops to mix the whole envelope in the early phase 
of outbursts. The solar composition is assumed for the envelopes 
otherwise specified, which 
is a good approximation for weak shell flashes.  
\citet{pri95} showed that the composition of ejecta 
is close to the solar for high mass accretion rates.

The above approximations may not be accurate enough to follow nova outbursts 
but sufficient for our purpose of qualitative study on the occurrence of winds 
and on the internal structures of WD envelopes.

\section{NOVA SEQUENCES} \label{sec_novaevolution}

\begin{figure}
\epsscale{1.05}
\plotone{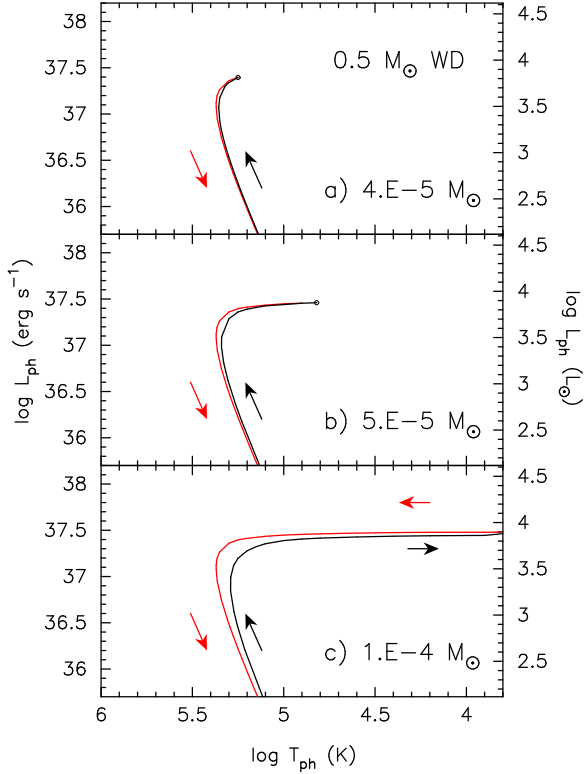}
\caption{Rising (black) and decay (red) phases of a nova outburst 
on a $0.5~M_\odot$ WD. The position of maximum expansion is indicated by 
a small open circle.  (a) $\Delta M_{\rm ig}=4 \times 10^{-5}~M_\odot$,
 (b) $5 \times 10^{-5}~M_\odot$, and (c) $1 \times 10^{-4}~M_\odot$. 
\label{novaM05}
}
\end{figure}

\begin{figure}
\epsscale{1.05}
\plotone{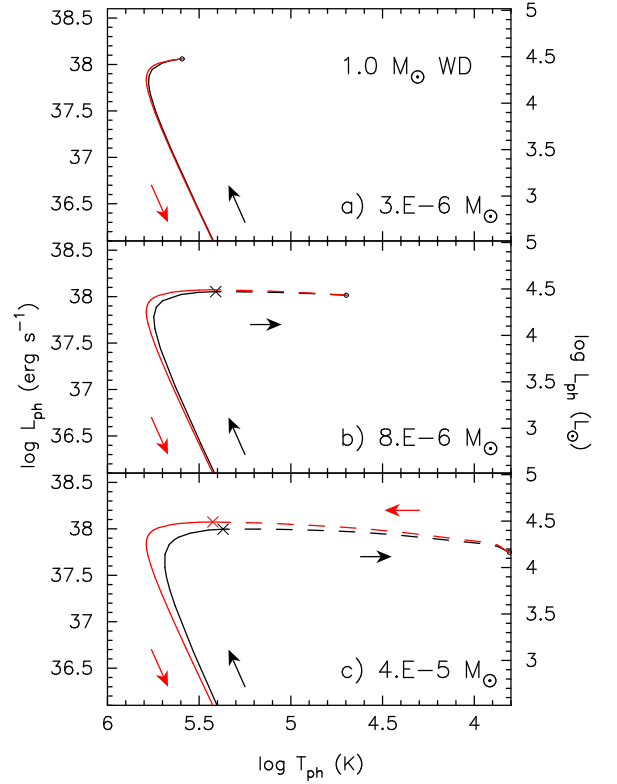}
\caption{Same as Fig.\ref{novaM05}, but for a $1.0~M_\odot$ WD.
 (a) $\Delta M_{\rm ig}=3 \times 10^{-6}~M_\odot$,  (b)
$8 \times 10^{-6}~M_\odot$, and  (c) $4 \times 10^{-5}~M_\odot$. 
The solid and dashed lines indicate static and wind phases, respectively. 
The position where the wind occurs/stops is denoted by a cross. 
\label{novaM10}
}
\end{figure}

\begin{figure}
\epsscale{1.05}
\plotone{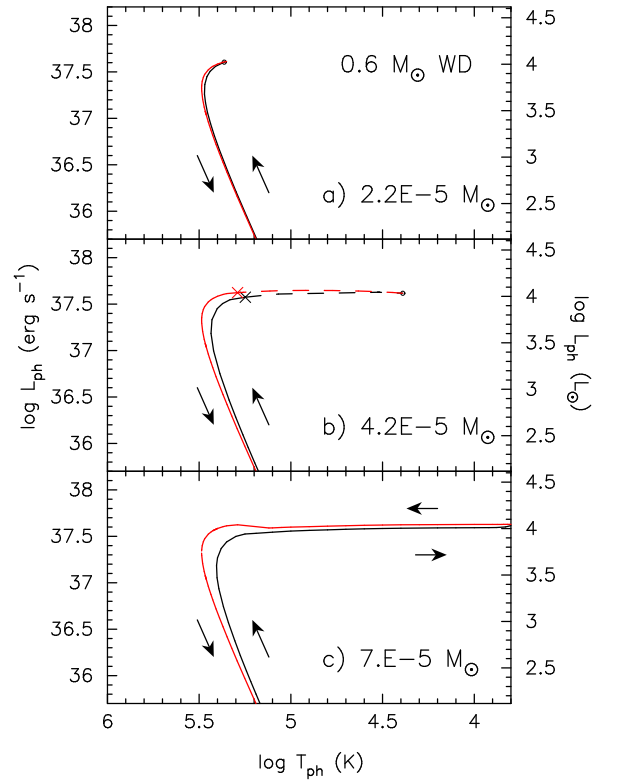}
\caption{Same as Fig.\ref{novaM05}, but for a $0.6~M_\odot$ WD.
 (a) $\Delta M_{\rm ig}=2.2 \times 10^{-5}~M_\odot$,  (b)
$4.2 \times 10^{-5}~M_\odot$, and  (c) $7 \times 10^{-5}~M_\odot$. 
\label{novaM06}
}
\end{figure}

Figures \ref{novaM05}, \ref{novaM10}, and \ref{novaM06} show the quasi-evolution 
sequences that mimic the rising and decay phases of nova outbursts. 
In a case of $0.5~M_{\odot}$ WD, no optically thick 
winds are accelerated at all, thus all the sequences in Figures \ref{novaM05}
consist of static solutions. Figure \ref{novaM05}a shows a case with   
a small ignition mass of $\Delta M_{\rm ig}=4.0 \times 10^{-5}~M_{\odot}$. 
The rising phase ends when the photospheric 
temperature and radius reach $\log T $ (K)$= 5.25$ and 
$\log r $ (cm)$=9.77~(0.085~R_\odot)$, respectively. This point is the 
maximum expansion where the envelope reaches thermal equilibrium.
The maximum expansion shifts toward lower temperature when we 
increase the ignition mass. In a case of $\Delta M_{\rm ig}=5.0 \times 10^{-5} 
~M_{\odot}$ (Fig. \ref{novaM05}b)  the temperature and radius at the 
maximum expansion 
are $\log T $ (K)$= 4.8$ 
and $\log r $ (cm)$=10.70~(0.73~R_\odot)$, respectively,
and for $\Delta M_{\rm ig}=1. \times 10^{-4}~M_{\odot}$ (Fig. \ref{novaM05}c), 
$\log T $ (K)$ = 3.79$ and $\log r $ (cm)$=12.71~(74~R_\odot)$, respectively. 

Figure \ref{novaM10} shows a $1.0~M_{\odot}$ WD case.
If we increase the ignition mass as $\Delta M_{\rm ig}=3 \times 10^{-6}, 
~8 \times 10^{-6}~M_{\odot}$, and  $4 \times 10^{-5}~M_{\odot}$, 
the temperature and radius at the maximum expansion change to 
$\log T $ (K)$= 5.6,~4.7$, and 3.8, and $\log r $ (cm)$=9.4~(0.036 R_\odot)$, 
~$11.18~ (2.2 R_\odot)$, and $12.83~(97 R_\odot)$, respectively.
We found that optically thick winds occur for $\Delta M_{\rm ig} > 3.6 
\times 10^{-6}~M_{\odot}$. 
We define this critical mass for winds as $\Delta M_{\rm wind}$, 
i.e.,  winds occur when $\Delta M_{\rm ig} > \Delta M_{\rm wind}$. 
As a wind is accelerated due to the
opacity peak at $\log T $ (K)$\sim 5.2$, optically thick winds occur in 
the lower temperature side of the peak \citep{kat94h} as shown by the dashed 
line in Figures \ref{novaM10}b and \ref{novaM10}c. 
If we further increase the ignition mass to $\Delta M_{\rm ig} > 8.5 \times 
10^{-4}~M_{\odot}$, winds are suppressed and static solutions exist 
instead of wind solutions. 
This is because hydrostatic balance is established in the envelope 
when a substantial amount of matter above the super-Eddington region, 
i.e., around the opacity peak, supresses down the wind-acceleration. 
We define this critical mass for static expansion,  
as $\Delta M_{\rm exp}$. If $\Delta M_{\rm ig} > \Delta M_{\rm exp}$ no winds 
are accelerated and the envelope expands without optically thick winds. 
However, this critical value is too large for the ignition mass of an 
accreting $1.0~M_{\odot}$ WD \citep{pri95}, so such static solutions are not 
realized and we don't go into details. 

These two critical masses of $\Delta M_{\rm wind}$ and $\Delta M_{\rm exp}$ 
depend on the WD mass. For a 0.6~$M_{\odot}$ WD, 
we obtained $\Delta M_{\rm wind}=2.3\times 10^{-5}~M_{\odot}$ and 
$\Delta M_{\rm exp}=6.5\times 10^{-5}~M_{\odot}$. Figure \ref{novaM06}a shows 
a case of $\Delta M_{\rm ig}=2.2 \times 10^{-5}~M_{\odot}$, in which 
no winds occur because 
the photospheric temperature does not reach the OPAL peak even at the 
maximum expansion at $\log T $ (K)= 5.35 and $\log r $ (cm)$=9.68~(0.069 R_\odot)$. 
In the case of $\Delta M_{\rm ig}=4.2 \times 10^{-5}~M_{\odot}$ (Fig. \ref{novaM06}b)  
the optically thick winds begin when the photospheric temperature decreases 
to $\log T$(K)$ = 5.25$. 
The maximum expansion reaches the photospheric radius of $\log r$(cm)$ =11.60~(
5.7 R_\odot)$ and the temperature of $\log T $(K) $= 4.39$. 
Figure \ref{novaM06}c shows the case of $\Delta M_{\rm ig}=
7.0 \times 10^{-5}~M_{\odot}$ ($> \Delta M_{\rm exp}$) in which the whole period 
of the outburst can be followed by a sequence of static solutions. 

Note that there is only one path for the decay phase of the $0.5~M_{\odot}$ 
WD when it is in a thermal equilibrium. The two decay phases in Figures 
\ref{novaM05}a and \ref{novaM05}b are a part of that in Figure \ref{novaM05}c. 
Similarly the decay phases of Figures \ref{novaM10}a and \ref{novaM10}b are 
identical to that of Figure \ref{novaM10}c. 
On the other hand, there are two different paths in the $0.6~M_{\odot}$ WD 
even when it is in thermal equilibrium. The decay phase of Figure \ref{novaM06}b 
consists of wind and static solutions, only the latter part of which is identical to 
that of Figure \ref{novaM06}c. The wind sequence in Figure \ref{novaM06}b 
and the static sequence at $\log T_{\rm ph} < 5.29$ in Figure \ref{novaM06}c 
are different from each other all in envelope structure, in 
evolution time scale, and in light curve 
as we will see later.

\section{INTERNAL STRUCTURES} \label{sec_internal}

\begin{figure}
\epsscale{1.0}
\plotone{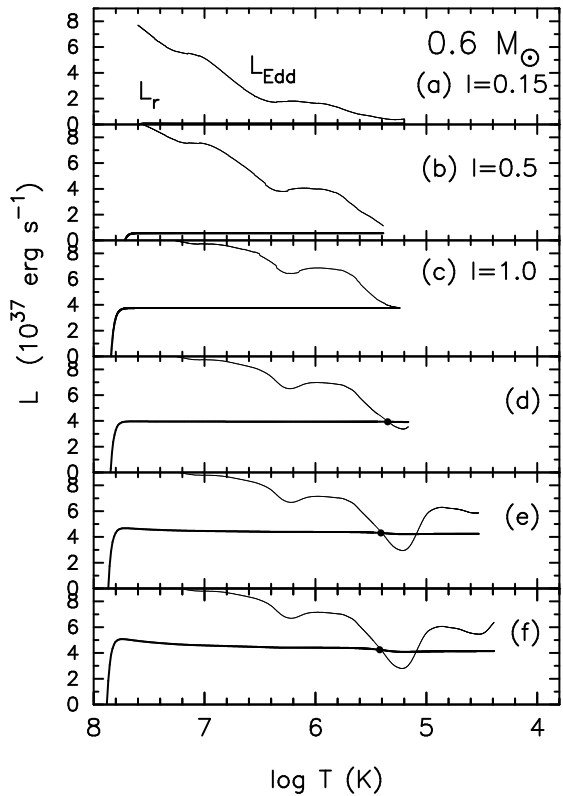}
\caption{ Evolutional change of the diffusive and Eddington luminosities 
in the rising phase of Figure \ref{novaM06}b (the envelope mass  
of $\Delta M_{\rm ig}=4.2 \times 10^{-5}~M_\odot$ 
on the 0.6 $~M_{\odot}$ WD).
The top three panels show the static solutions for different $l$, the ratio of 
the luminosity to the Eddington luminosity at the photosphere. 
The bottom three panels show the optically thick wind solutions in which 
the small dot denotes the critical point \citep{kat83a}. The right edge 
of each curve corresponds to the photosphere, the radius of which 
is  (a) $\log R_{\rm ph}$~(cm)=9.08,  (b) 9.23,
 (c) 9.86,  (d) 10.05, (e) 11.33, and  (f) 11.60. The model (a) locates 
under the ``knee'' of the line in Fig. \ref{novaM06}b so the photospheric 
radius and temperature are smaller than those of model (b). 
\label{LLedT_dM423E-5} 
}
\end{figure}

Figure \ref{LLedT_dM423E-5} shows the distribution of the 
diffusive luminosity and the local Eddington luminosity against the 
temperature for solutions along the rising phase of Figure \ref{novaM06}b, i.e.,
$0.6~M_\odot$ WD of $\Delta M_{\rm ig}=4.2 \times 10^{-5}~M_{\odot}$.
Here, the local Eddington luminosity is defined as

\begin{equation}
L_{\rm Edd} \equiv {4\pi cGM \over\kappa},
\end{equation}

\noindent
where $\kappa$ is the opacity in which we use the OPAL opacity \citep{igl96}. 
Since the opacity $\kappa$ is a function of the temperature and density, 
the Eddington luminosity is also a local variable. 
This Eddington luminosity has a local minimum at $\log T$ (K) = 5.25 
corresponding to the opacity peak as shown in  Figure \ref{LLedT_dM423E-5}f.

As the star moves upward in Figure \ref{novaM06}b, the diffusive luminosity 
increases and approaches the Eddington luminosity near the photosphere 
(Figures \ref{LLedT_dM423E-5}a,b, and c). 
When the photospheric temperature decreases to $\log T$ (K) $\sim 5.2$, 
matter is accelerated and steady mass-loss begins (Figure 
\ref{LLedT_dM423E-5}d).
After that, the envelope continuously expands to until point C in Figure 
\ref{hr_diagram}. 
A narrow super-Eddington region appears corresponding to the opacity peak 
at $\log T$ (K) $\sim 5.2$.

\begin{figure}
\epsscale{1.05}
\plotone{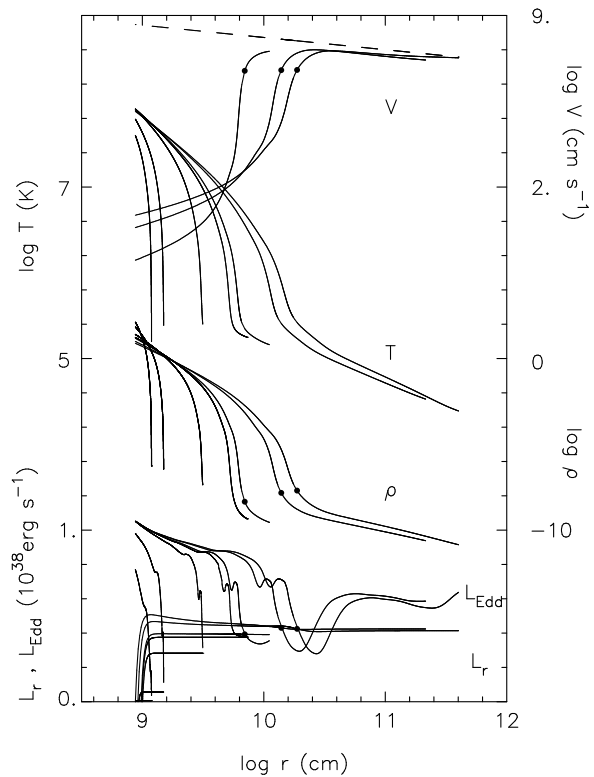}
\caption{Evolutional change of internal structure of the envelope in the rising phase of 
Figure \ref{novaM06}b (0.6 ~$M_\odot$ WD with $\Delta M_{\rm ig}=4.2\times 10^{-5}~
~M_\odot$). From top to bottom, the wind velocity
($V$), temperature ($T$), density ($\rho$), Eddington luminosity 
($L_{\rm Edd}$) and diffusive luminosity ($L_{\rm r}$).
The dashed line denotes the escape velocity defined by 
$\sqrt{2GM_{\rm WD}/r}$. The right edge of each curve 
corresponds to the photosphere. 
Dot denotes the critical point of each wind solution.
\label{internal_dM423E-5}
}
\end{figure}

The change of envelope structure is shown in Figure \ref{internal_dM423E-5}.
We see that the structure of the static 
solution just before the wind occurs is very similar to that 
of the adjacent wind solution. This property has been already pointed out by 
\citet{kat85} for the old opacity and here we confirm it for the OPAL opacity
having a prominent peak at a much higher temperature region.

\begin{figure}
\epsscale{1.05}
\plotone{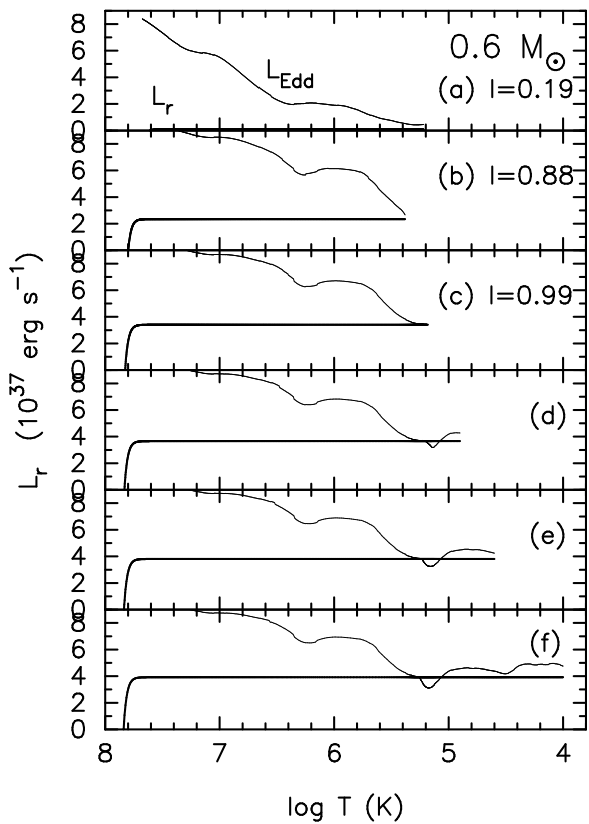}
\caption{Same as  Fig. \ref{LLedT_dM423E-5}, but for the rising phase of 
Fig. \ref{novaM06}c (0.6~ $M_{\odot}$ WD with $\Delta M_{\rm ig}=7.0 
\times 10^{-5}~M_\odot$). 
The photospheric radius of each solution is  (a) $\log R$~(cm)=9.11, (b)
9.50, (c) 9.96, (d) 10.56, (e) 11.16, and (f) 12.37. 
\label{LLedT_dM7E-5}
}
\end{figure}

\begin{figure}
\epsscale{1.05}
\plotone{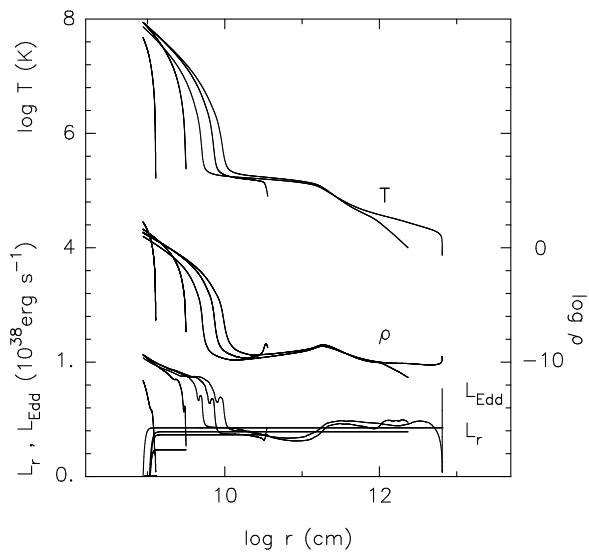}
\caption{Same as Fig. \ref{internal_dM423E-5}, but for the rising phase of
Fig. \ref{novaM06}c (0.6 ~$M_\odot$ WD with $\Delta M_{\rm ig}=7.0 \times 10^{-5}
~M_\odot$). 
\label{internal_dM7E-5}
}
\end{figure}

In the massive envelope of $\Delta M_{\rm ig}=7.0 \times 10^{-5}~M_{\odot}$, 
however, no winds arise when the envelope expands beyond the opacity peak
of $\log T$ (K) $\sim 5.2$.    
Figures \ref{LLedT_dM7E-5} and \ref{internal_dM7E-5} show internal 
structures in such an evolution sequence. Difference from 
the wind sequence  becomes prominent as the 
envelope expands (see Figs. \ref{LLedT_dM423E-5} and \ref{internal_dM423E-5}).

A remarkable difference can be observed in the density distribution.  
Figure \ref{internal_dM423E-5} shows the 
monotonically decreasing density having a $r^{-2}$ dependence in the outer envelope 
($\log r$ (cm) $\geq 10.3$) as expected from the equation of continuity 
$4\pi r^2\rho v =$ const, where the velocity $v$ is almost constant 
in an outer part of the envelope. 
On the other hand, the envelope in the static sequence (Fig. 
\ref{internal_dM7E-5})  
develops large density-inversion at $\log r$ (cm) $\sim 10-11.3$  
corresponding to a super Eddington region. 
This density-inversion arises in order to keep hydrostatic balance 
in the super-Eddington region ($L_{\rm Edd} < L_r$) as expected 
from the equation of hydrostatic balance. 
Inefficient convections occur in the region of  $L_{\rm Edd} < L_r$ but are 
unable to carry all of the diffusive energy flux, thus the structure is 
super-adiabatic. Comparing Figure 
\ref{internal_dM7E-5} with Figure \ref{internal_dM423E-5}
we see a prominent core-halo structure in the density and temperature distributions
in the static solutions.

\begin{figure}
\epsscale{1.15}
\plotone{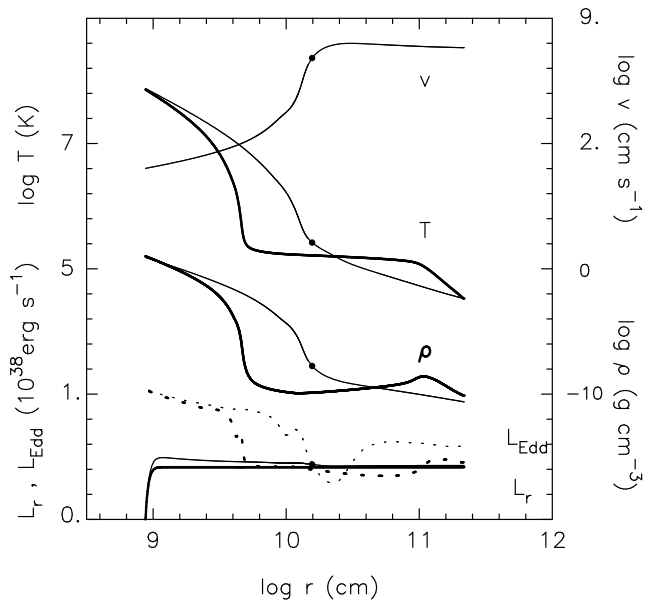}
\caption{Comparison of two solutions with the same photospheric 
temperature of $\log T_{\rm ph} = 4.53$ in the decay phase of 
0.6 $~M_{\odot}$ WD. {\it Thick line}: envelope solution 
in the static sequence. The envelope mass is 
$2.3 \times 10^{-5}~M_{\odot}$. {\it Thin line}: envelope solution 
in the wind sequence. The envelope mass is $3.8 \times 10^{-5}~M_{\odot}$. 
The right edge of each line corresponds to the photosphere. 
\label{internal_decay}
}
\end{figure}

Such differences between the wind and static sequences 
can be also seen in the decay phase. 
Figure \ref{internal_decay} demonstrates the difference between the two solutions 
in the decay phase of $0.6~M_{\odot}$ WD. 
Both the static and wind solutions have the same photospheric temperature 
$\log T_{\rm ph}$ (K) $= 4.53$
but they are very different in their internal structures. The mass-losing envelope 
shows the density distribution of $r^{-2}$ in the outer part  
while the quasi-static envelope develops a wide density-inversion region. 

\section{OCCURRENCE OF OPTICALLY THICK WINDS} \label{sec_variousWDmass}

As we have seen in the previous sections, the wind mass-loss 
occurs in a limited range of the envelope mass. 
When the ignition mass is smaller than $\Delta M_{\rm wind}$ the envelope 
expands a little and the photospheric temperature 
does not reach the opacity peak of $\log T$ (K) $\sim 5.2$
(see Figs. \ref{novaM10}a and \ref{novaM06}a).
Therefore, no optically thick winds occur. On the other hand, 
if the ignition mass is larger than $\Delta M_{\rm exp}$, 
winds are suppressed in a way that density-inversion 
balances to radiation-pressure gradient in a super-adiabatic region 
(see Fig. \ref{internal_dM7E-5}). In this case the envelope expands without 
optically thick winds. 
Therefore, optically thick winds occur only for 
$\Delta M_{\rm wind} < \Delta M_{\rm ig} < \Delta M_{\rm exp}$ as we have 
already seen in the 0.6~$M_\odot$ WD.

\begin{figure}
\epsscale{1.05}
\plotone{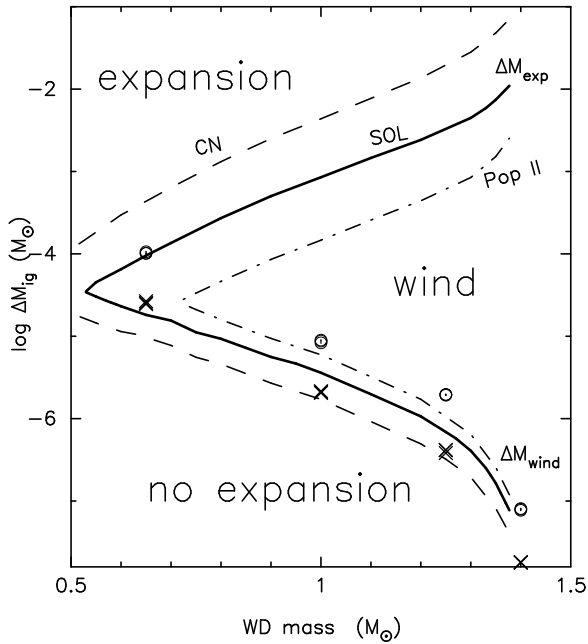}
\caption{Upper and lower critical ignition masses for winds, 
$\Delta M_{\rm exp}$ and  $\Delta M_{\rm wind}$, plotted against the 
WD mass. The optically thick winds occur 
in the right hand side of the lines (labeled ``wind''). 
In the upper side of the line, optically thick 
wind does not occur and the envelope expands quasi-statically  
(labeled "expansion").
In the lower region, the envelope does not expand so much ("no expansion") and
no wind arises. {\it Thick line}: solar composition. {\it Dashed line}:
CNO rich $(X=0.35,~Y=0.33$,~$X_{\rm CNO}=0.30$,and $Z=0.02)$.  
{\it Dash-Dotted line}: Population II composition $(X=0.7, Z=0.004)$. 
Open circles with a dot and crosses denote mass ejection model 
and no mass ejection model \citep{pri95}, respectively. 
See text for more details. 
\label{lsmaxdM}
}
\end{figure}

Figure \ref{lsmaxdM} depicts these two critical masses of $\Delta M_{\rm wind}$ 
and $\Delta M_{\rm exp}$ for various WD masses. Optically thick 
winds are driven in the right hand side of the thick solid line. 
In the lower region $(\Delta M_{\rm ig} < \Delta M_{\rm wind})$
the envelope does not expand much and no optically thick winds arise.
In the upper region $(\Delta M_{\rm ig} > \Delta M_{\rm exp}$) the envelope is 
so massive and winds are suppressed. 
Here, we label the lower region ``no expansion'', and the upper region 
``expansion'', which is an abbreviation 
of ``quasi-static expansion''. 

As the OPAL opacity depends on the chemical composition of the envelope, 
both $\Delta M_{\rm exp}$ and $\Delta M_{\rm wind}$ depend on the composition. 
For a composition of typical classical novae ($X=0.35, Y=0.33, X_{\rm CNO}
=0.30,$ and $Z=0.02)$  the wind is strongly 
accelerated and the ``wind'' region extends as shown in Figure \ref{lsmaxdM}.
On the other hand, for Population II stars with lower metal content, the wind 
is weak \citep{kat97,kat99} and the ``wind'' region becomes narrower.

Figure \ref{lsmaxdM} also shows that $\Delta M_{\rm exp}$ increases with the 
WD mass and reaches as large as $10^{-3}- 10^{-2}~M_{\odot}$ for $1.0-1.38 
~M_{\odot}$ WDs. Such a large ignition mass is unlikely to be 
realized in accreting WDs. 
Dynamical calculations have shown that the envelope
mass is up to $2\times 10^{-4}~ M_{\odot}$ for a 1.0 ~$M_{\odot}$ WD 
and $3\times 10^{-6}~M_{\odot}$ for a 1.4 $~M_{\odot}$ for the 
accretion rate of $10^{-12.3}~M_{\odot}$~yr$^{-1}$ and 
$10^{-11}~M_{\odot}$~yr$^{-1}$ for cold WDs \citep{yar05}. 
These values are practically upper limits for the envelope mass 
in nova outbursts but are still much smaller than $\Delta M_{\rm exp}$. 
Therefore, we regard that the "expansion" region and a part of the upper 
``wind'' region theoretically exist but may not be realized 
in the actual mass-accreting $1.0-1.38~M_\odot$ WDs.

In less massive WDs such as $\sim 0.6~M_{\odot}$, on the other hand, 
$\Delta M_{\rm exp}$ is as small as $10^{-4}~M_{\odot}$, which 
corresponds to the mass-accretion rates of $\sim 10^{-8}~M_{\odot}$~yr$^{-1}$ 
\citep{pri95}, a typical mass accretion rate 
in cataclysmic variables. Therefore, the ``expansion'' region becomes 
a subject of realistic interest. 
If the ignition mass is larger than $\Delta M_{\rm exp}$ the envelope 
expands in a quasi-static manner. 
In less massive WDs of $M_{\rm WD} < 0.5~M_{\odot}$ the ``wind'' region disappears 
completely and the envelope evolves in a quasi-static manner.

\section{DISCUSSION} \label{sec_discussion}

\subsection{Light Curves in the Decay Phase}

\begin{figure}
\epsscale{1.05}
\plotone{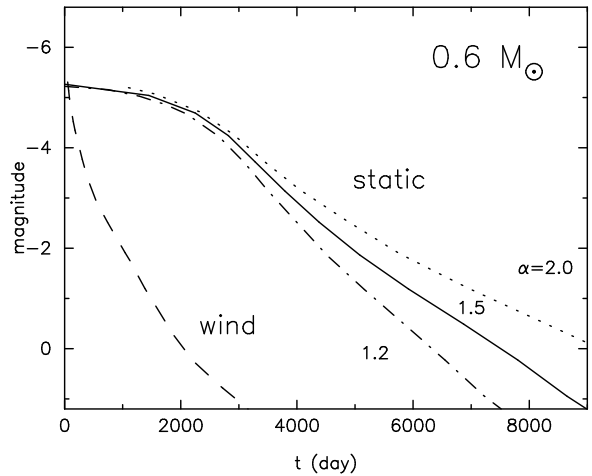}
\caption{Comparison of light curves in the decay phase of a 0.6~ $M_{\odot}$ 
WD between the "wind" sequences (dashed) and "static" expansion sequences (solid) 
for the same chemical composition.
{\it Dotted line}: Static expansion sequence with mixing length-parameter 
$\alpha=2.0$. {\it Solid line}: $\alpha=1.5$. 
{\it Dash-dotted line}: $\alpha=1.2$. 
}
\label{lightM06}
\end{figure}

As we see in Figure \ref{novaM06} there are two different sequences 
that represent the decay phase of nova outbursts on a 0.6~$M_\odot$ WD.
These two sequences are different from each other in its envelope mass and 
mass decreasing rate, so the light curves are also different.
Figure \ref{lightM06} shows the light curves corresponding to these sequences, 
both of which start from $\log T_{\rm ph}$ (K) $\sim 3.9$.  
The light curve of the static sequence decays much slowly, because its 
evolution speed is determined by the mass-decreasing rate due only to 
hydrogen nuclear burning, whereas in the "wind" sequence the evolution is 
accelerated by the winds which carry away a large part of the envelope matter.
The most remarkable difference is in the flat peak of the static 
sequence, whereas it decays sharply in the wind sequence.  

The flat peak in the static sequence reminds us a peculiar light curve of the 
symbiotic nova PU Vul, which shows a very long plateau peak of 3000 days
followed by a slow decline of 3 magnitudes/2500 days. The 
spectra suggest a very quiet expansion with no indication
of wind mass loss during the flat peak \citep{iij84,kan91,kol95,yam82}. 
We may apply our static "expansion" sequences to such slow outbursts.
Detailed light curve fitting will be presented separately.

Both of the wind and static expansion solutions are certainly stable 
because they represent nova envelope and red giant envelope, respectively.   
However, it is interesting to point out a possibility that a nova evolves along the 
static expansion sequence but it suddenly jumps in the wind sequence or vice versa  
during the course of evolution. If this kind of transition occurs it 
may proceed from a higher total enthalpy state to a lower one.
The enthalpy integrated for the entire envelope is, for example, $3.6 
\times 10^{52}$~erg for a static envelope solution of 
$\Delta M =5.4 \times 10^{-4}~M_\odot$ 
with $\log T_{\rm ph}$ (K) $=3.8$, and $1.0 \times 10^{52}$~erg for a wind mass-loss 
solution of a similar envelope mass and photospheric temperature.
Therefore, if the transition occurs it possibly goes from the static 
expansion sequence to the wind sequence. 
As the internal structure of the two solutions are very different (see
Fig. \ref{internal_decay}) and the excess energy will be released, 
the transition may not occur in a quiet way 
but accompany some violent activities. It may be interesting to connect such 
activities to oscillatory behaviors often observed  
in early light curves of slow/symbiotic novae.

Figure \ref{lightM06} demonstrates that both of 
the static "expansion" and "wind" sequences exist for the 
same WD mass and the same chemical composition of the envelope. 
If the accretion rate onto the WD changes with time and, as a 
result, the ignition mass changes from one outburst to the next around 
$\Delta M_{\rm exp}$, these two different types of outbursts can be realized 
on the same WD.  
In such a case outbursts behave very differently for a small change 
of mass-accretion rate. 

In this way, we expect many active phenomena associated with static 
expansion sequences. More quantitative studies including light curve analysis 
of slow/symbiotic novae are necessary.

\subsection{Comparison with Dynamical Calculations}\label{sec_comparison}

We have obtained $\Delta M_{\rm wind}$, the lower critical mass for 
the envelope having optically thick winds.
This critical mass is compared with hydrodynamical calculations by \citet{pri95}
who presented a number of shell flashes for various parameters. 
The open circles with a dot in Figure \ref{lsmaxdM} denote their models in which 
mass ejection occurs, and the crosses indicate the models with no mass ejection.
In these models the chemical composition of the envelope is not uniform but 
close to the solar value because dredge-up of WD material is insignificant 
for high mass-accretion rates.
Therefore, these models can be compared with our model labeled ``solar''. 
Our critical line  is quite consistent with their results except the
$0.65~M_{\odot}$ WD, in which their crosses are above our line.
These authors defined the "mass ejection" by that 
the expansion velocity exceeds the escape velocity, which is different from 
our definition of occurrence condition of the winds  described in 
\S \ref{sec_internal}. 
In our wind solutions of $0.6~M_\odot$ the velocity does not exceeds 
the escape velocity as shown in Figure \ref{internal_dM423E-5}. 
Considering this difference and other difference of input parameters, 
we can say that our $\Delta M_{\rm wind}$ is quite consistent with results 
in \citet{pri95}.

\subsection{Dependence on the Mixing Length Parameter} \label{sec_mixinglength}

We assumed a mixing-length parameter of $\alpha=1.5$.
Many authors have estimated the $\alpha$ parameter 
to be $\alpha=1.2 - 2.0$ \citep[][and references therein]{asi00,pal02}   
for various types of stars. 
The mixing-length parameter could be a function of stellar 
parameters or could depend on stellar structure, but 
these dependences are not known yet. 
Therefore, we simply adopted $\alpha=1.5$. 

In order to see the dependence of light curves on 
the mixing-length parameter we have calculated additional models   
with $\alpha=1.2$ and  $2.0$ as shown in Figure \ref{lightM06}. 
For a smaller $\alpha$, the energy transport is more efficient and the  
star evolves faster, resulting in a steeper light curve. 
Even though, the static sequences still show a long-lasted flat peak 
and slow decline after that.

\section{SUMMARY} \label{sec_summary}

Our main results are summarized as follows;

(1) For a given WD mass and chemical composition, the optically thick wind 
occurs when the ignition mass ($\Delta M_{\rm ig}$) satisfies the condition, 
i.e., $\Delta M_{\rm wind} < \Delta M_{\rm ig} < \Delta M_{\rm exp}$.
If $\Delta M_{\rm ig} < \Delta M_{\rm wind}$ the envelope does not expand and 
no wind is accelerated. When 
$\Delta M_{\rm ig} > \Delta M_{\rm exp}$, winds are suppressed and the 
envelope expands with no optically thick winds.

(2) Optically thick winds occur smoothly from a static envelope because 
the structure is not drastically changed before and after the wind occurs. 
This property was already reported in \citet{kat85} for the old opacity but 
we confirm this for the OPAL opacity.   

(3) For a given WD mass and chemical composition, there exist two solutions 
with different structures. One is the wind solution 
with a monotonic decrease in the density distribution. Another is the  
static solution that develops a wide super-adiabatic region with density inversion.

(4) In massive WDs ($\geq 0.8~M_\odot$), $\Delta M_{\rm exp}$ is very large 
and the static-expansion sequence 
is not practically realized ($\Delta M_{\rm ig} \ll \Delta M_{\rm exp}$).
In the intermediate mass WDs ($0.6-0.7~M_\odot$), both the optically 
thick wind and quasi-static expansion are realized depending on 
the ignition mass.
In less massive WDs ($\leq 0.5~M_\odot$) no optically thick wind occurs 
independently of the ignition mass.

(5) The wind sequence has been applied to nova outbursts in which strong 
mass-loss is observed. The newly found quasi-static expansion sequence may be 
applied to slow evolutions of symbiotic novae like PU Vul, which shows 
a long-lasted flat peak with no indication of strong winds.




\acknowledgments

This research has been supported in part by the Grant-in-Aid for
Scientific Research (20540227)
of the Japan Society for the Promotion of Science.


\begin{thebibliography}{}



\bibitem[Asida (2000)]{asi00} Asida, S.M. 2000, \apj, 528, 896


\bibitem[Fujimoto (1982)]{fuj82} Fujimoto, M.Y. 1982, \apj, 257, 752,

\bibitem[Iben (1982)]{ibe82}
Iben, I. Jr. 1982, \apj, 259, 244

\bibitem[Iglesias \& Rogers (1996)]{igl96}
Iglesias, C. A., \& Rogers, F. J. 1996, \apj, 464, 943

\bibitem[Iijima \& Ortolani (1984)]{iij84} Iijima,T., \& Ortolani,S.
1984, \aap, 136, 1

\bibitem[Kanamitsu et al. (1991)]{kan91} Kanamitsu, O., Yamashita, Y., 
Norimoto, Y., Watanabe, E., Yutani, M.  1991, \pasj, 43, 523


\bibitem[Kato (1983a)]{kat83a}
Kato, M. 1983a, \pasj,  35, 33

\bibitem[Kato (1983b)]{kat83b}
Kato, M. 1983b, \pasj,  35, 507

\bibitem[Kato (1985)]{kat85}
Kato, M. 1985, \pasj,  37, 19

\bibitem[Kato \& Hachisu (1988)]{kat88} Kato, M., \& Hachisu, I.
1988, \apj, 329, 808


\bibitem[Kato (1997)]{kat97}
Kato, M. 1997, \apjs, 113, 121

\bibitem[Kato (1999)]{kat99}
Kato, M. 1999, \pasj, 51, 525


\bibitem[Kato \& Hachisu (1989)]{kat89}
Kato, M., \& Hachisu, I., 1989, \apj, 346, 424

\bibitem[Kato \& Hachisu (1994)]{kat94h}
Kato, M., \& Hachisu, I., 1994, \apj, 437, 802

\bibitem[Kato et al.(1989) Kato, Saio, \& Hachisu]{kat89sh}
Kato, M., Saio, H., \& Hachisu, I., 1989, \apj, 340 509

\bibitem[Kolotilov et al.(1995) Kolotilov, Munar \& Yudin]{kol95}
Kolotilov,E.A., Munari,U., \& Yudin,B.F., 1995, \mnras 275,185

\bibitem[MacDonald (1983)]{mac83} MacDonald, J. 1983, \apj, 267,732

\bibitem[Nariai et al.(1980) Nariai, Nomoto, \& Sugimoto]{nar80}
Nariai, K., Nomoto, K., \& Sugimoto, D. 1980, \pasj, 32, 473


\bibitem[Paczy\'nski \& \.Zytkow (1978)]{pac78}
Paczy\'nski, B. \& \.Zytkow,  A.N. 1978, \apj, 222, 604

\bibitem[Palmieri et al. (2002)]{pal02} Palmieri, R., Piotto, G., 
Saviane, I., Girardi, L.  \& Castellani,V. 2002, \aap, 392, 115


\bibitem[Prialnik et al. (1986)]{pri86}
Prialnik, D., 1986, \apj, 310,222


\bibitem[Prialnik \& Kovetz (1995)]{pri95}
Prialnik, D., \& Kovetz, A. 1995, \apj, 445,789

\bibitem[Sion et al.(1979) Sion, Acierno, \& Tomczyk]{sio79}
Sion, E. M., Acierno, M.J., \& Tomczyk, S.
1979, \apj, 230, 832

\bibitem[Sparks et al. (1978)]{spa78}
Sparks, W. N., Starrfield, S., \& Truran J. W. 1978, \apj, 220, 1063

\bibitem[Yaron et al.(2005)]{yar05} Yaron, O., Prialnik, D., Shara, M.M., 
Kovetz, A. 2005, \apj, 623, 398


\bibitem[Yamashita et al. (1982) Yamashita, Maehara \& Norimoto]{yam82}
Yamashita, Y., Maehara, H., Norimoto, Y., 1982, \pasj, 34,269

\end{thebibliography}
\end{document}